\documentclass[epj]{webofc}
\usepackage[varg]{txfonts}   
\usepackage{natbib}
\def\pt{\ensuremath{p_{\mathrm{T}}}\xspace}
\def\mus{\ensuremath{\mu}s\xspace}

\woctitle{Connecting The Dots / Intelligent Trackers 2017}
\begin{document}
\title{FPGA-Based Tracklet Approach to Level-1 Track Finding at CMS for the HL-LHC}

\author{
\firstname{Edward} \lastname{Bartz}\inst{2} \and
\firstname{Jorge} \lastname{Chaves}\inst{1} \and
\firstname{Yuri} \lastname{Gershtein}\inst{2} \and
\firstname{Eva} \lastname{Halkiadakis}\inst{2} \and
\firstname{Michael} \lastname{Hildreth}\inst{3} \and
\firstname{Savvas} \lastname{Kyriacou}\inst{2} \and
\firstname{Kevin} \lastname{Lannon}\inst{3} \and
\firstname{Anthony} \lastname{Lefeld}\inst{4} \and
\firstname{Anders} \lastname{Ryd}\inst{1} \and
\firstname{Louise} \lastname{Skinnari}\inst{1} \and
\firstname{Robert} \lastname{Stone}\inst{2} \and
\firstname{Charles} \lastname{Strohman}\inst{1} \and
\firstname{Zhengcheng} \lastname{Tao}\inst{1} \and
\firstname{Brian} \lastname{Winer}\inst{4} \and
\firstname{Peter} \lastname{Wittich}\inst{1} \and
\firstname{Margaret} \lastname{Zientek}\inst{1}\fnsep\thanks{\email{mez34@cornell.edu}} 
}

\institute{Cornell University, Ithaca, NY, USA
\and
Rutgers University, New Brunswick, NJ, USA
\and 
University of Notre Dame, South Bend, IN, USA
\and
Ohio State University, Columbus, OH, USA}

\abstract{%
During the High Luminosity LHC,
the CMS detector will need charged particle tracking at the hardware trigger level to maintain a manageable trigger rate and achieve its physics goals. 
The tracklet approach is a track-finding algorithm based on a road-search algorithm that has been implemented on commercially available FPGA technology. 
The tracklet algorithm has achieved high performance in track-finding and completes tracking within 3.4 \mus on a Xilinx Virtex-7 FPGA. 
An overview of the algorithm and its implementation on an FPGA is given, 
results are shown from a demonstrator test stand and system performance studies are presented. 
}
\maketitle
\section{Introduction}
\label{sec:intro}

CERN's LHC accelerator complex will undergo major upgrades, planned for 2025, to increase the 
instantaneous luminosity up to around $7.5 \times 10^{34}$ $\mathrm{cm}^{-2}\mathrm{s}^{-1}$ \cite{ref:hl-lhc}. 
This era of the High Luminosity LHC (HL-LHC) will yield an average number of overlapping proton-proton collisions per bunch crossing (pileup or PU) between 140 and 200. 
The LHC collides proton bunches every 25 ns. 
If every collision were kept, this would mean roughly 20 to 40 Tbps would need to be stored.
However, only a small fraction of these collisions are of interest for further study.  
The CMS \cite{ref:cms} physics program for the HL-LHC requires being able to intelligently select or trigger on collisions that could contain interesting physics scenarios. 
The trigger system reduces the 40 MHz collision rate to a manageable data storage rate of 400 Hz. 

\begin{figure}[t!]
\centering
\includegraphics[width=7cm]{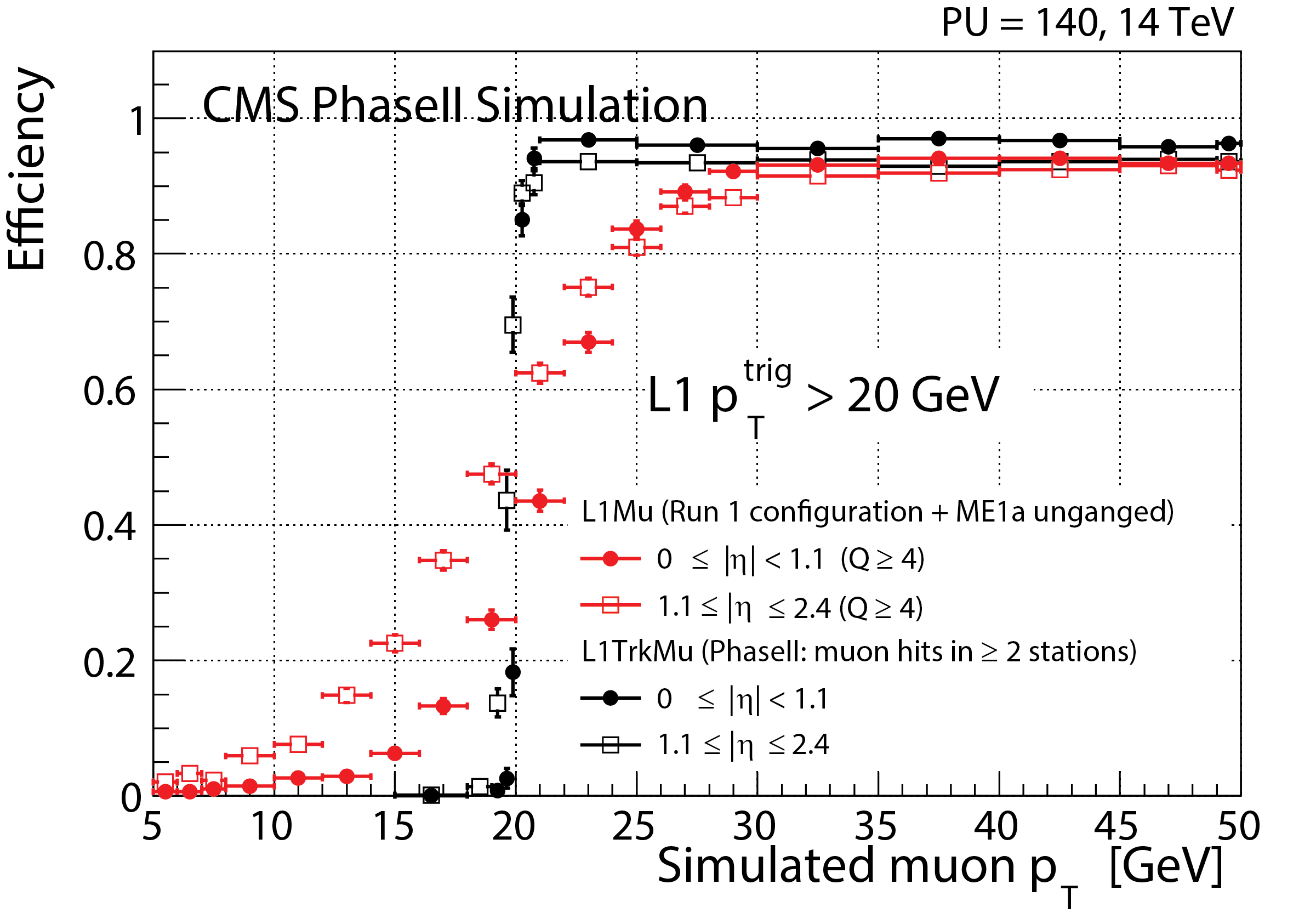}
\includegraphics[width=7cm]{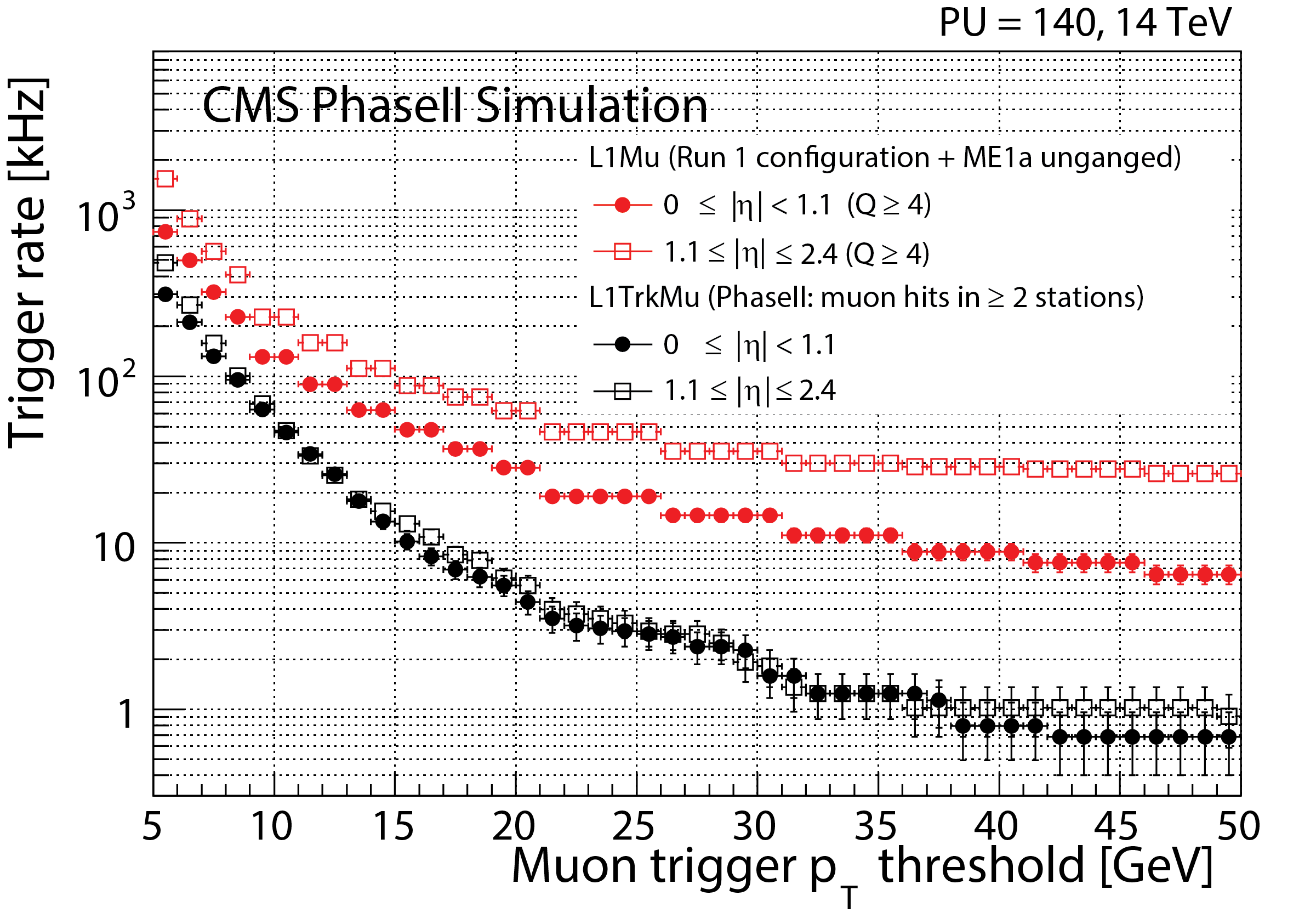}
\caption{ The efficiency of a single muon trigger with \pt $ > 20$ GeV as a function of \pt(left) and the trigger rate as a function of the muon trigger threshold (right), shown for the stand-alone muon trigger (red) and when including L1 tracking from the silicon tracker (black) for various $\eta$ ranges \cite{ref:tech-proposal}.}
\label{fig:HL-LHC_L1}
\end{figure}

An integral part of reducing the rate is the hardware Level-1 (L1) trigger system. 
At the HL-LHC, the L1 trigger rates for objects such as single muons, electrons, or jets will exceed the current front-end capabilities. 
Increasing the transverse momentum (\pt) trigger thresholds for these objects limits the physics potential of the HL-LHC and no longer sufficiently reduces the data rate. 
Integrating charged particle tracking from the silicon tracker into the L1 trigger will improve lepton identification and momentum measurements and provide track isolation and vertex identification. 
These additional handles have the potential to reduce the L1 rates while maintaining reasonable trigger thresholds. 
One example of this is shown in Figure~\ref{fig:HL-LHC_L1}, where efficiency (left) and rate (right) are plotted for a single muon trigger. 
The red curves show the behavior of a stand-alone L1 muon trigger system, while the black curves show the performance of the same triggers when including L1 tracking. 
In particular, adding L1 tracking improves the momentum measurement which translates to a sharp turn-on curve at the trigger threshold and a reduced trigger rate \cite{ref:tech-proposal}. The space available to store each event in the buffer limits the time allowed for the trigger decision. 
The L1 trigger decision must occur within 12~\mus. In order to leave time for correlating tracks with other physics objects and the trigger decision, 
the track reconstruction must be made in approximately 4 \mus.

The upgraded (``Phase 2'') CMS detector will include an all silicon tracker that is designed to integrate tracking into the L1 trigger decision. 
The new tracker consists of a pixel detector, which is not used in L1 tracking, and an outer tracker that has a central barrel and two endcaps. 
The outer tracker will have the unique feature of ``\pt modules'' that provide \pt discrimination at the level of the front-end readout electronics. 
In the detector's magnetic field, a charged particle's bend (and consequent hit pattern) depends on its \pt. 
Correlating hits between two closely spaced sensors thus provides a coarse \pt measurement. 
A \pt module consists of two sensors, either instrumented with pixels or strips, that are mounted 1--5 mm apart.
Two different modules are used: pixel-strip (PS) modules in the inner layers of the barrel and inner half of the endcaps, 
and strip-strip (2S) modules in the outer layers of the barrel and outer half of the endcaps. 
Further details on the layout and module design can be found in Ref. \cite{ref:tech-proposal}. 
The pixelated sensor of the PS modules provides a precise measurement of the displacement along the beam line ($z$-axis), which enables primary vertex reconstruction at the L1 level. 
Correlated hits (``stubs") in \pt modules are required to be consistent with a \pt $> 2$ GeV track originating from the interaction point. 
Since most minimum bias events have low \pt tracks, this provides a factor of $\approx 10$\% data reduction already. 
The stubs are then sent to the L1 track finding system. 

To summarize, the goals of the L1 tracking system are to reconstruct the trajectories of charged particles that have \pt $>$ 2 GeV and 
identify the $z_0$ of the track ($z$ coordinate where the track intercepts the $z$-axis) with about 1 mm precision, similar to the average separation of vertices in events with an average PU = 140. 
Additionally, the L1 track finding must be completed within 4 \mus. 
There are three possible CMS implementations to L1 tracking: (i) the ``tracklet'' approach, 
a road-search algorithm implemented using commercially available FPGAs and  the approach that is presented here, 
(ii) a Hough-transform based approach also using FPGAs \cite{ref:tmtt}, 
and (iii) an associative memory based approach using a custom ASIC \cite{ref:AM}.
The next sections will detail the tracklet algorithm, implementation on hardware, performance results, and some projections towards making a full system for completing L1 tracking at CMS at the HL-LHC.  

\section{Tracklet Algorithm}
\label{sec:aglo}

\begin{figure}[ht]
\centering
\includegraphics[width=4.5cm]{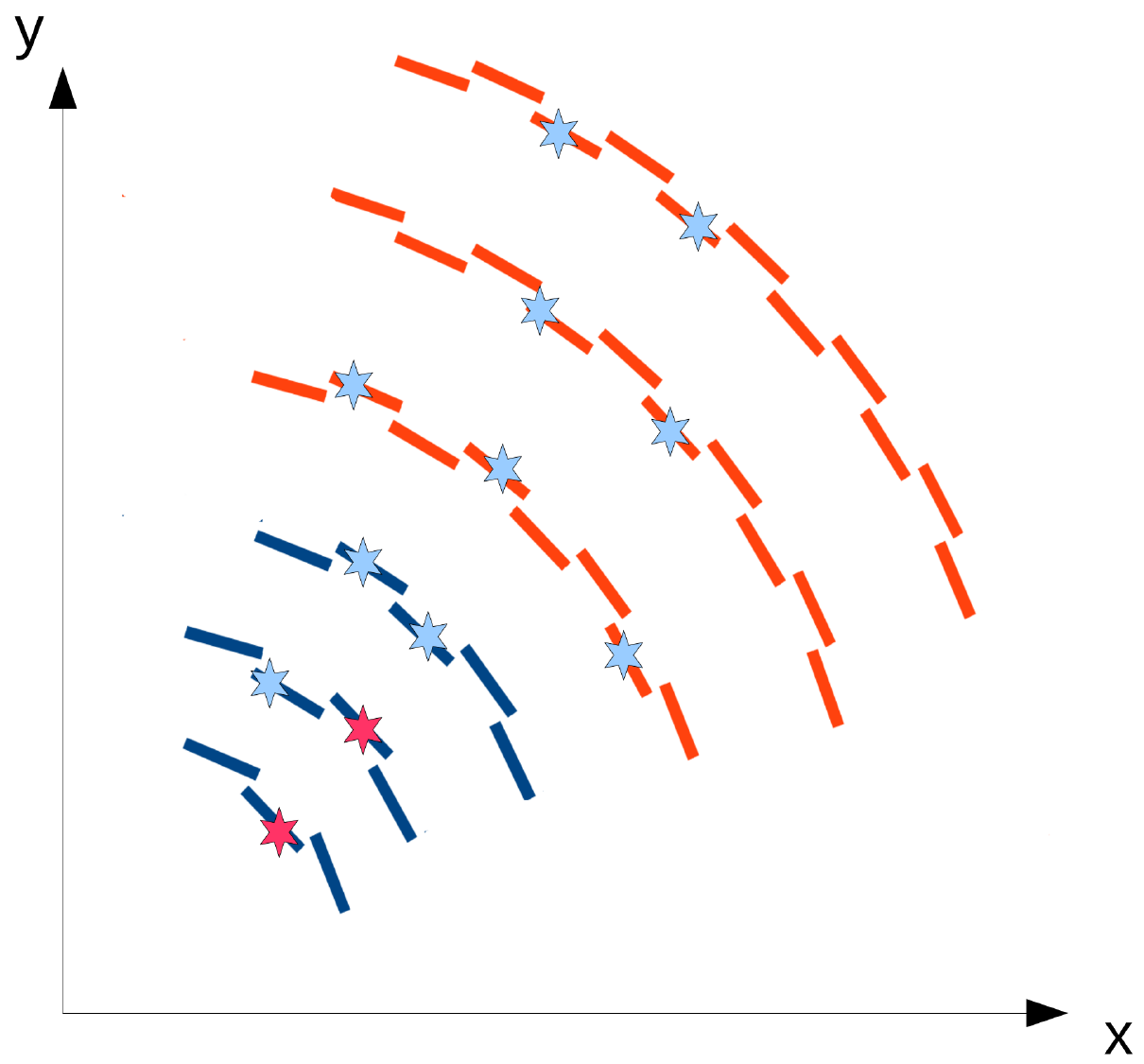}
\includegraphics[width=4.5cm]{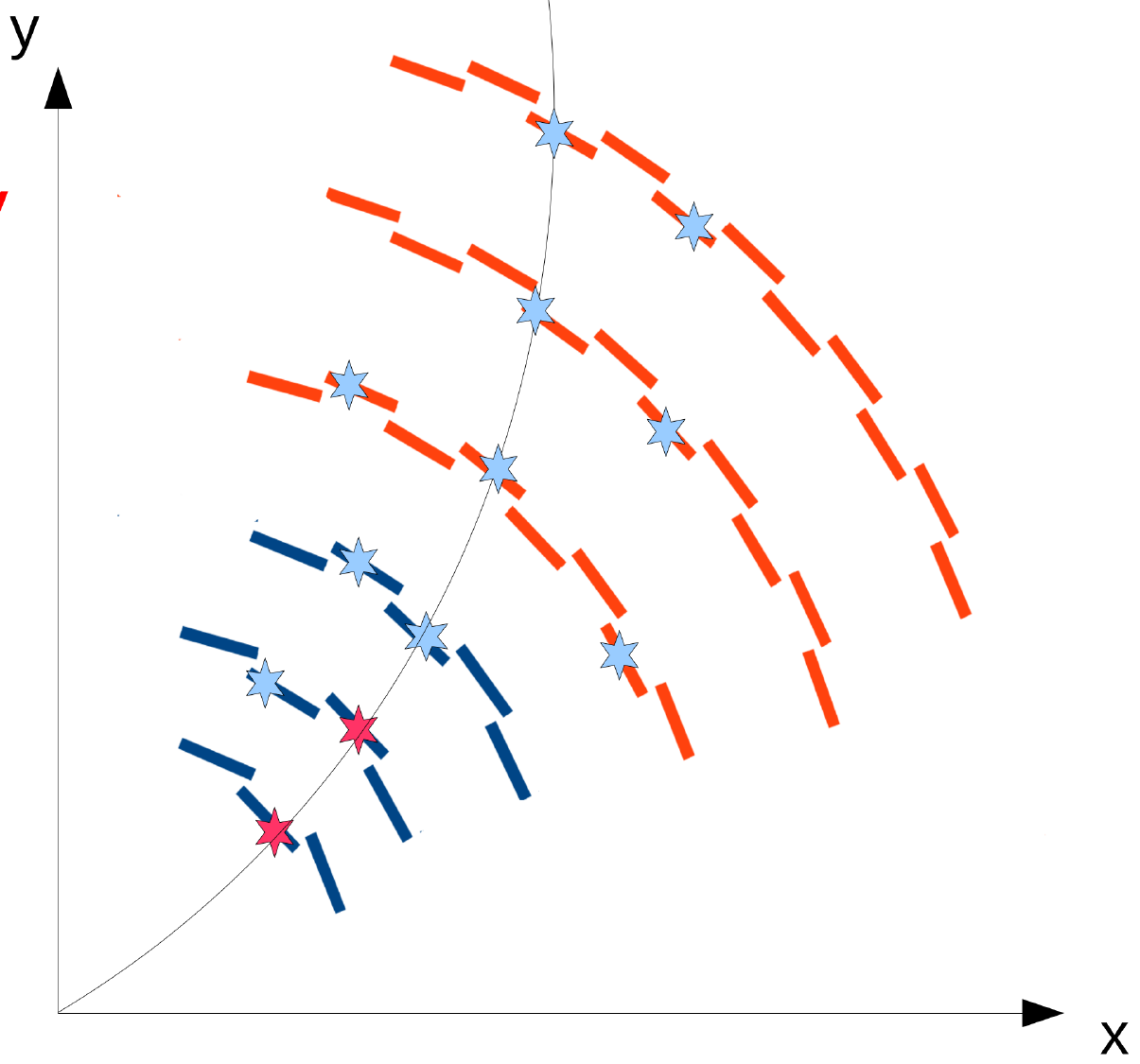}
\includegraphics[width=4.5cm]{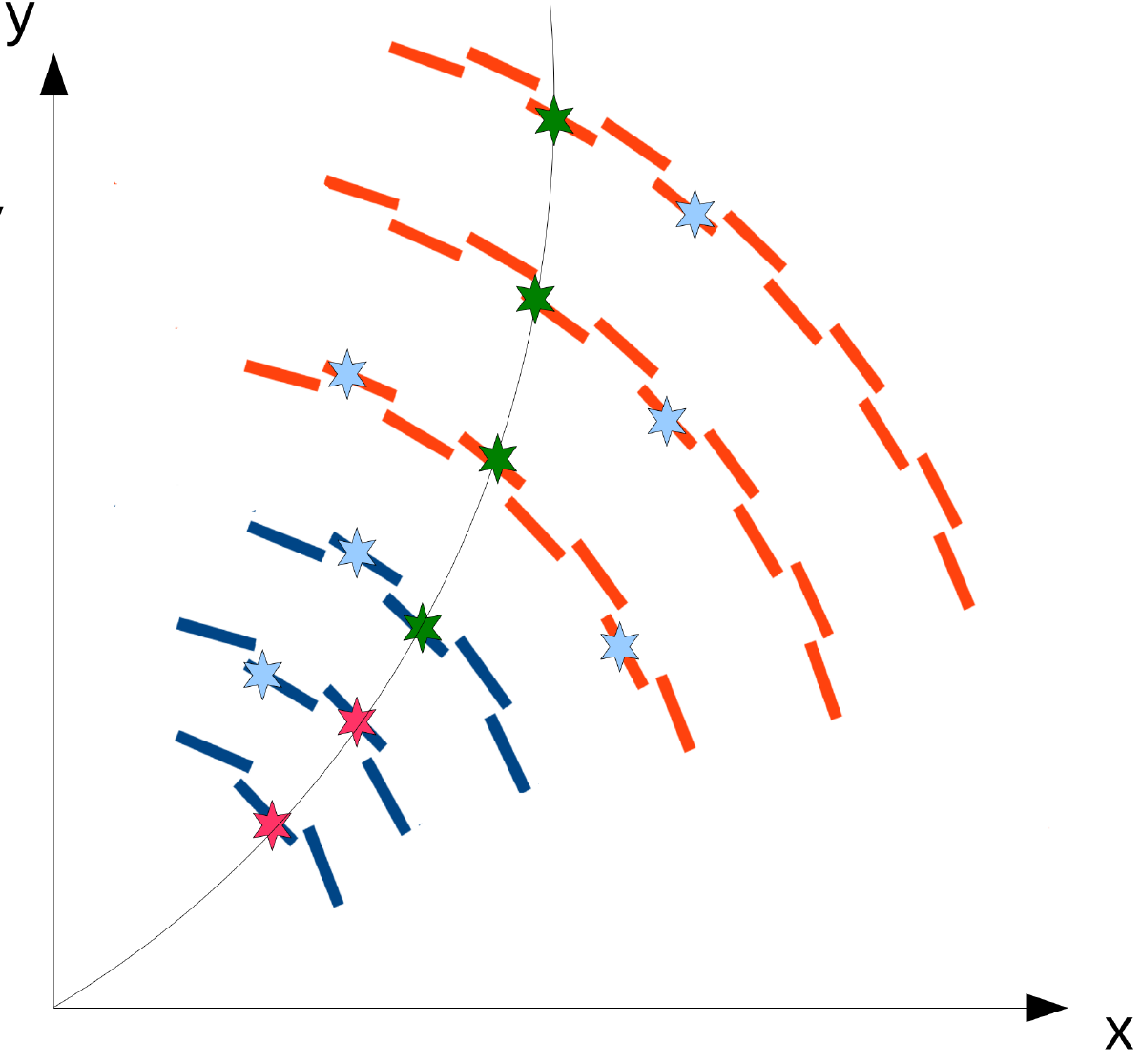}
\caption{
In the first step (left) of the algorithm a pair of stubs (red) in adjacent layers are combined to form a tracklet. 
The trajectory of the tracklet is projected (middle) to the other layers. 
Stubs in the other layers that are close to the projection (green) are selected as matches (right) to the tracklet. Final track parameters are calculated using all associated stubs. 
}
\label{fig:algosketch}
\end{figure}

\begin{figure}[ht]
\centering
\includegraphics[width=10cm]{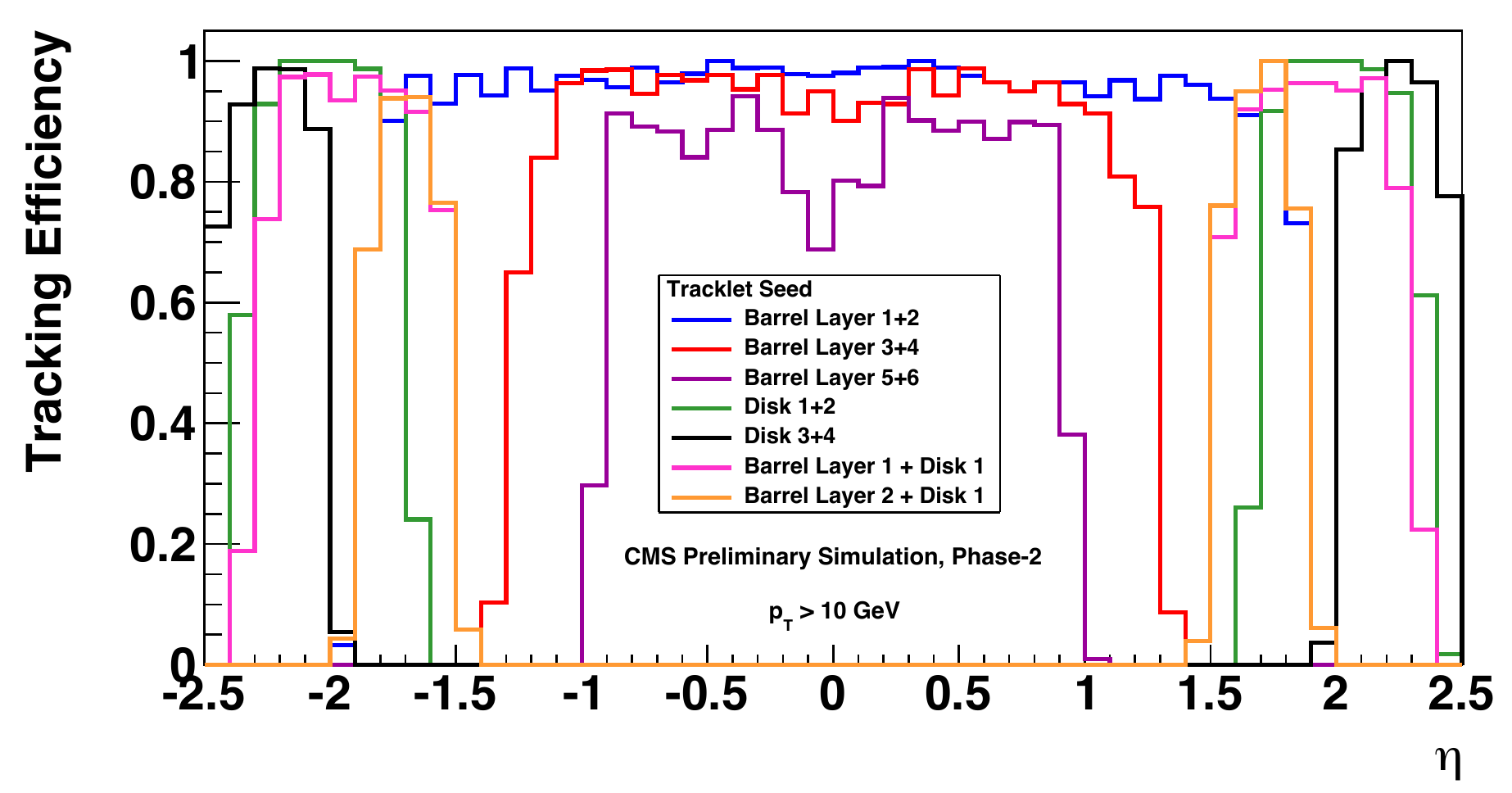}
\caption{Efficiency for finding a single muon track for several different seeding layer/disk combinations as a function of $\eta$. Here it is seen that the combination of the different seeding pairs provides coverage of the full $\eta$ range of the detector \cite{ref:tracklet-louise}.}
\label{fig:seed_eff}
\end{figure}

A sketch of the algorithm steps, which are detailed below, is shown in Figure \ref{fig:algosketch}. 
The tracklet approach starts by forming track seeds (tracklets) from pairs of stubs in adjacent barrel layers or endcap disks. 
The tracklet is an initial estimate of the tracklet parameters calculated from these two stubs using the interaction point as a constraint. 
A candidate tracklet must be consistent with a \pt $>$ 2 GeV track that originates within $|z_0| < 15$ cm. 
The seeding is performed for several combinations to provide good coverage of the entire pseudorapidity ($\eta$) range of the detector. 
The tracking efficiency for different seeding combinations is shown for single muons in Figure \ref{fig:seed_eff}, 
using an integer-based C++ emulation of the algorithm as it would be implemented on an FPGA.
In the current implementation of the algorithm, seeding includes pairs between layers 1+2, 3+4, and 5+6, and between disks 1+2 and 3+4. 

The track parameters of the tracklets are then projected to other layers and disks to search for consistent stubs. 
When the tracklets are projected to other layers/disks, the search for matching stubs occurs in predetermined search windows, derived from residuals between projected tracklets and stubs. 
The projection of the tracklets occurs both inwards and outwards (i.e. to and from the interaction point). 
If a stub is found that is consistent with the original tracklet's parameters, the matched stub is included in the track candidate and the difference between the projected tracklet position and 
the matched stub position is stored.

A linearized $\chi^2$ fit is performed using all stubs in the track candidate - the stubs used to make the original tracklet plus the matched stubs.
The track fit uses pre-calculated derivatives and the projection-stub differences. 
The linearized $\chi^2$ fit corrects the initial tracklet parameters giving the final track parameters: \pt, $\eta$, $z_0$, the azimuthal angle at the closest approach ($\phi_0$),
and optionally the impact parameter of the transverse plane ($d_0$). 
Because seeding is performed for multiple seeding combinations, a single track may be found several times. 
Duplicated tracks are removed by comparing the found tracks in pairs, comparing the number of independent and shared stubs.

\section{Hardware System}
\label{sec:hardware}

To address the challenging amount of data and limited processing time available, the tracklet hardware configuration relies on massively parallelizing the data processing. 
The main parallelization is the division of the detector into sectors in the $r$-$\phi$ plane. 
The current project uses 28 $\phi$ sectors. 
This was chosen so that tracks with \pt $>$ 2 GeV span a maximum of two sectors.
This limits the need for data transfer between sectors to the nearest neighboring sector on each side. 
Tracklets that project to a neighboring sector are sent there for tracklet-stub matching. 
A dedicated processing board is used for each $\phi$ sector. 
A small amount of data is duplicated every other layer to avoid gaps in the track-finding. 

To allow for more time for data processing, the whole 28 $\phi$ sector system is replicated $n$ times using a round robin time multiplexing approach. 
Each independent time multiplexed system receives a new event every $n$ $\times$ 25 ns.  
The choice of $n$ is driven by a balance of cost, efficiency and needed processing power. 
For the full system, $n = 4-8$ are considered reasonable choices.
The current implementation assumes a time multiplexing factor of $n=6$, so a new event is received every 150 ns. 

\subsection{Hardware Implementation}
\label{sec:hw_impl}

The tracklet algorithm is implemented in the firmware as several processing steps (names in bold) \cite{ref:tracklet-jorge}:
\begin{itemize}
\item \emph{Stub organization}: (1) Sort input stubs by layer (\textbf{Layer Router}), and (2) into smaller units in $z$ and $\phi$ called ``virtual modules'' (\textbf{VM Router}).
\item \emph{Tracklet formation}: (3) Select possible pairs of stubs (\textbf{Tracklet Engine}), and (4) calculate the tracklet parameters (\textbf{Tracklet Calculator}).
\item \emph{Projection organization}: (5) Transmit projections that point to neighboring sectors (\textbf{Projection Transceiver}), and (6) route the projections based on their virtual module (\textbf{Projection Router}). 
\item \emph{Stub matching}: (7) Match projections to stubs (\textbf{Match Engine}), and (8) calculate the tracklet-stub difference in positions (\textbf{Match Calculator}), and (9) transmit matches back to original sector (\textbf{Match Transceiver}).
\item \emph{Track fit}: (10) Fit final tracks (\textbf{Track Fit}), and (11) remove duplicate tracks (\textbf{Duplicate Removal}).
\end{itemize}

The largest combinatoric challenges occur at tracklet formation and match finding.
With an average PU = 140, there are $\approx 60$ stubs per layer per $\phi$ sector, this would yield $\approx 3600$ candidate tracklets per seeding combination \cite{ref:tracklet-jorge}.
By dividing each $\phi$ sector into smaller virtual modules, the tracklet formation and match finding processes are further parallelized.
Furthermore, only a small fraction of virtual module pairs are consistent with \pt $>$ 2 GeV and $|z_0| < 15$ cm tracks. 
This reduces the number of stub combinations that need to be tried at each of these steps. 
Consequently, the number of tracklets per $\phi$ sector is reduced to $\approx 20$ per seeding combination. 

All of the processing steps above read from memories filled by the previous step and write the output to another set of memories.
Currently all processing steps are synchronized to a single common 240 MHz clock. 
By construction, the system is fully pipelined and operates at a fixed latency. 
When a new event arrives (currently every 150 ns) the previous event moves to the next processing step. 
This necessarily implies that a given step can only perform a fixed number of operations.
If the time limit is reached, processing on the remainder of the data for that step stops, meaning any remaining data must be truncated. 
The effect of truncation on the system is minimal, and more details on the performance with truncation are presented later. 

\subsection{Hardware Demonstrator}
\label{sec:hw_demo}

\begin{figure}[ht]
\centering
\includegraphics[width=10cm]{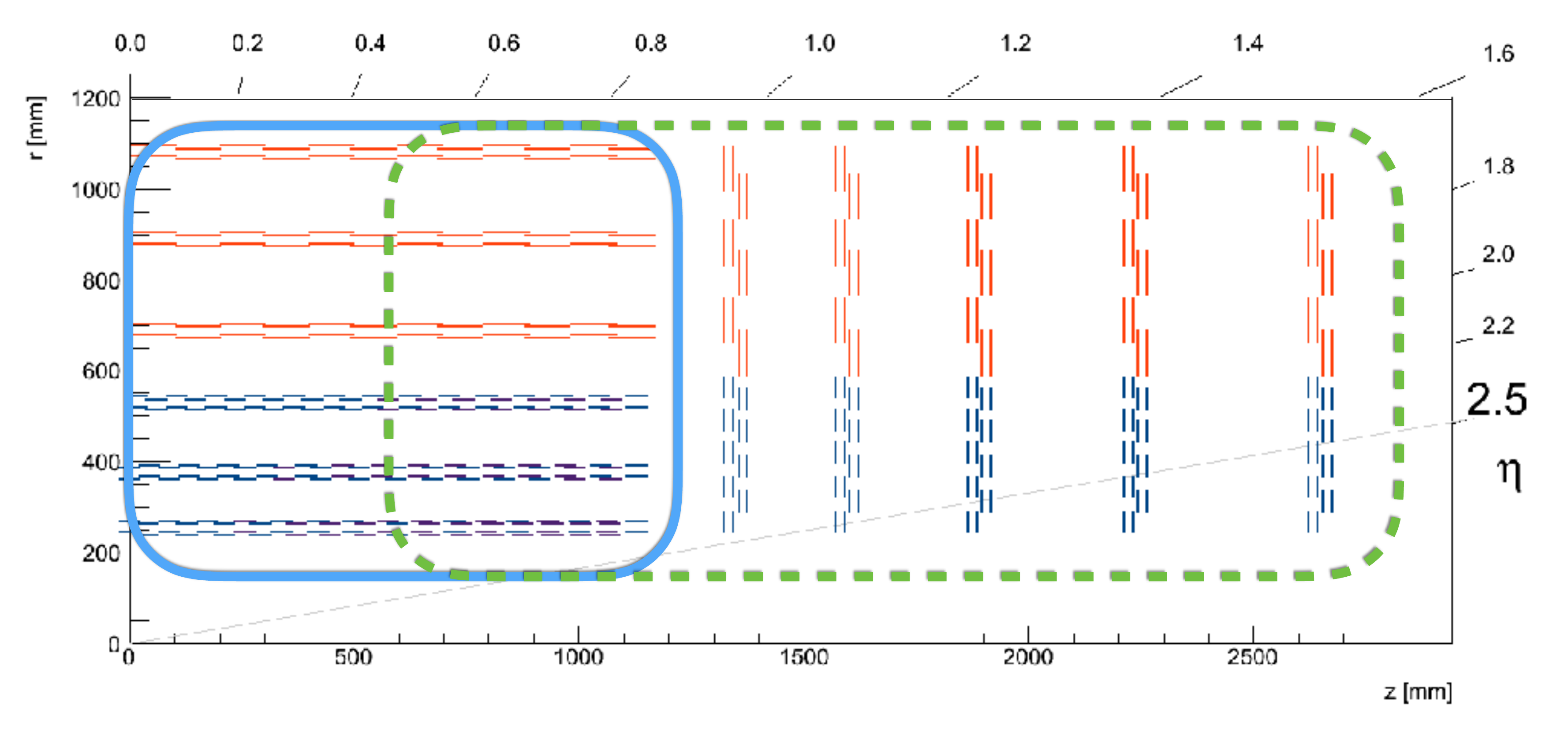}
\caption{Sketch of the $+z$ portion of the detector in the $r$--$z$ plane. PS modules are shown in blue while 2S modules are shown in red. 
The box in blue shows the detector region covered by the half barrel implementation.
A box in green shows the detector region covered by the other project that spans a quarter of the barrel, the transition region between the barrel and endcaps, and the endcaps.}
\label{fig:proj_sketch}
\end{figure}

The full tracklet algorithm, including all processing and transmission steps, has been implemented in firmware. 
Two complete implementations - one for half the barrel ($+z$) and one for a quarter of the barrel plus the forward endcaps - were used to demonstrate the feasibility of this approach for the full $\eta$ range of the detector. 
A sketch of the upgraded CMS tracker region covered by the each of the implementations is shown in Figure \ref{fig:proj_sketch}.
Note, the layout of the upgraded tracker will actually have tilted modules in the inner barrel layers with $\eta > 0.6$. The porting of the tracklet algorithm to the updated geometry is almost complete.

\begin{figure}[t!]
\centering
\includegraphics[width=8cm]{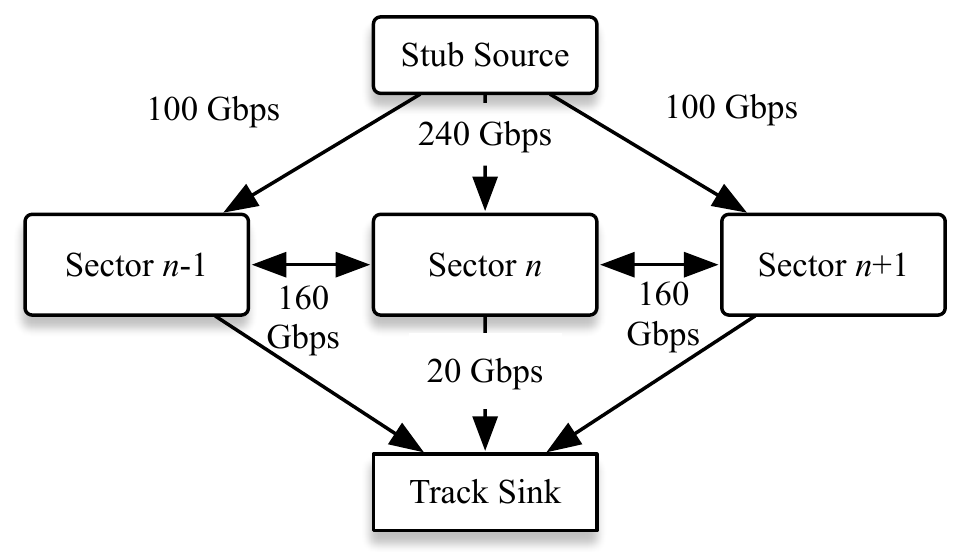}
\includegraphics[width=5.5cm]{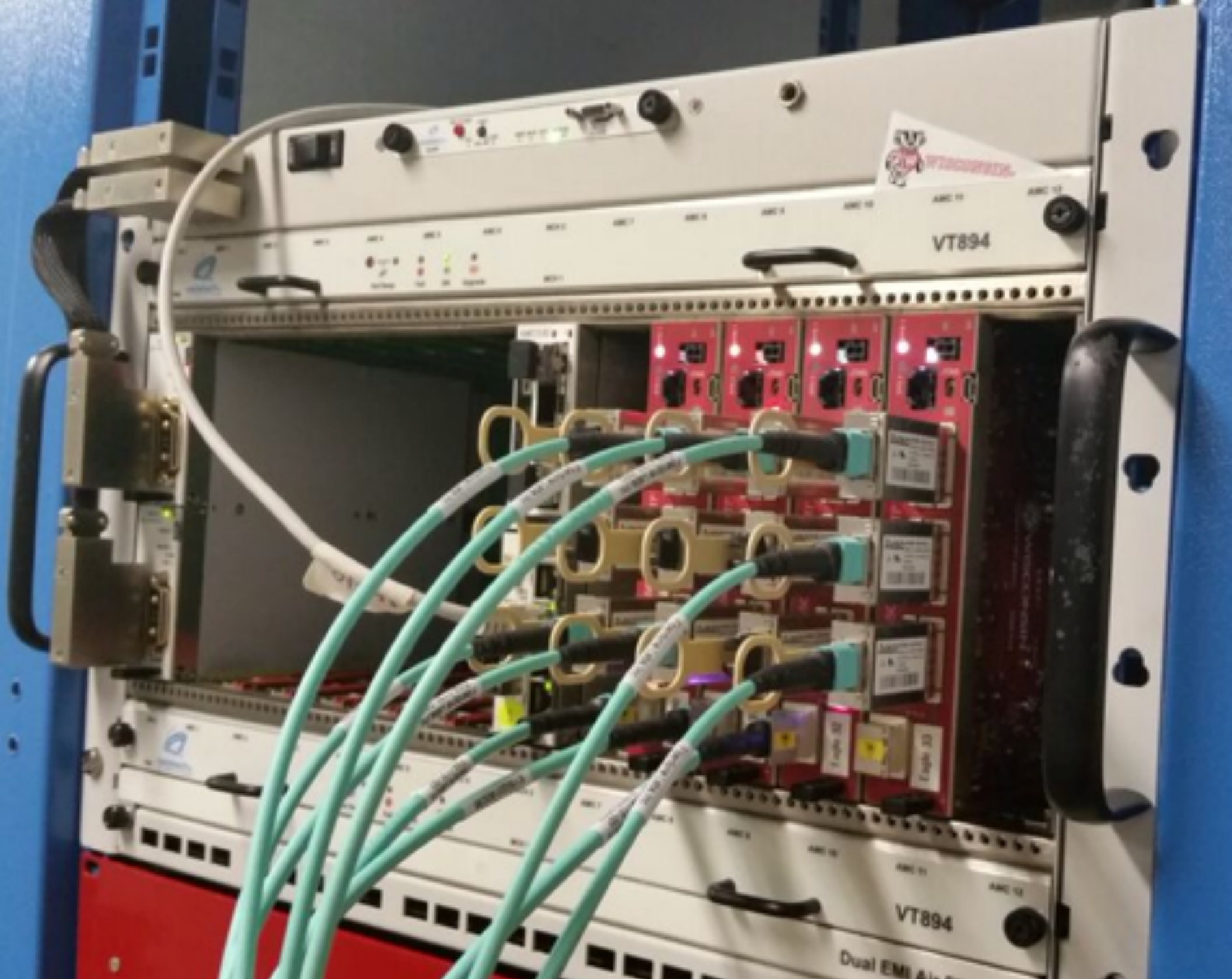}
\caption{Schematic overview (left) and image (right) of the tracklet demonstrator system. The stub source and track sink are on the same processing board. Estimated data transfer rates between the boards are also shown.}
\label{fig:demo_sketch}
\end{figure}

A system hardware demonstrator was set up for full scale testing of the firmware implementation. 
The demonstrator was used to show that the full L1 tracking chain meets the required performance within the available latency. 
The demonstrator test stand is one slice of the $n=6$ time multiplexed system. 
It includes three $\phi$ sector processing boards: one for the central $\phi$ sector, and one for each of its nearest neighbors. 
One additional processing board has the duplicate function of sending the stubs into the $\phi$ sector boards and receiving the final track outputs. 
A schematic of the demonstrator is shown in Figure \ref{fig:demo_sketch}. The central $\phi$ sector processor is the actual system under test.

In the final system, each sector processor is foreseen to be an ATCA blade with a Virtex Ultrascale+ FPGA. 
The current demonstrator system instead is made of four $\mu$TCA blades, called CTP7 boards \cite{ref:ctp7}. 
Each CTP7 has a Xilinx Virtex-7 (XC7VX690T) FPGA \cite{ref:virtex7} and a Xilinx Zynq-7000 SoC processor for configuration and outside communication. 
The CTP7 boards were developed for the current CMS L1 trigger \cite{ref:L1TDR}. 
An AMC13 \cite{ref:amc13} card provides synchronization between the boards with a central clock distribution. 
The inter-board communication uses 8b/10b encoding with 10 Gbps links. The demonstrator system is shown in Figure \ref{fig:demo_sketch}. 

\subsection{Demonstrator System Latency}
\label{sec:hw_latency}

Each processing step outlined in Section \ref{sec:hw_impl} takes a fixed number of clock cycles to process its input data. 
Hence it is feasible to calculate a model of the latency for the complete system. 
Calculations are done assuming a 240 MHz clock and a time multiplexing factor of six, the current configuration of the project.  
The latency for each processing module to receive data and produce the first result varies between 1-50 clock cycles depending on the module. 
Each processing step continues to process data for the same event for 150 ns  before switching to the next event. 

For some of the steps, data has to be transmitted between boards. 
Tracklet projections and their corresponding matched stubs must be sent to the neighboring sector processors. 
In these cases, the latency due to inter-board communication and links is included in the latency model.  
The measured transmission latency is 316.7 ns (76 clock cycles). 
This latency includes all parts of the transmission: the transceiver TX and RX, channel bonding (the use of multiple serial links to send the data), data propagation through 15 m long optical fibers, and time needed to prepare and pass data from processing modules to transceivers.  
The latency also includes the data transmission latency for receiving stubs from and sending final tracks back to the data source/sink processing board. 

A summary of the estimated latency is shown in Table \ref{tab:latency}. 
The total estimated latency for receiving the first track from an event is 3345.8 ns.
Because of the fixed-processing time, the final processed track for any event will come within 150 ns of the first track. 
The total latency of the demonstrator has also been measured with a clock counter located on the data source/sink board. 
The measured first track out latency is 3333 ns, which agrees within three clock cycles (0.4\%) with the modeled latency. 
This is already below the goal of 4 \mus for L1 tracks.

There are notable places where the latency can be improved. 
First, the layer router has become redundant as the incoming stubs are now foreseen to be sorted by layer. 
Removing this processing module will reduce the latency by 150 ns.  
The transmission protocol contributes a large amount to the latency of the system. 
By sending duplicated stubs from the neighboring sectors to the central sector, the need for inter-board communication 
in the projection transceiver and match transceiver steps is removed. 
This could shave off as much as $\approx$ 1 \mus from the total latency. 
Additional optimization in transmission protocol and clock speed will provide speedup as well.

\begin{table}[tb]
  \caption{Demonstrator latency model. For each step,
    the processing time and latency is given. For steps involving data
    transfer, the link latency is given. The total model latency is 3345.8 ns.}
  \label{tab:latency}
  \centering 
  \begin{tabular}{llllll}
    \hline
    Step & Proc.    & Step          & Step         & Link   & Step  \\
         & time        & latency      & latency     & delay    & total \\ 
        & (ns)        & (CLK)        & (ns)           &  (ns)       & (ns) \\ \hline 
    Input link              &    0.0 &    1  &   4.2   &  316.7  &  320.8   \\  
    Layer Router            &  150.0 &    1  &    4.2   &   -  &  154.2   \\  
    VM Router               &  150.0 &    4  &   16.7   &   -  &  166.7        \\  
    Tracklet Engine         &  150.0 &    5  &   20.8   &   -  &  170.8        \\  
    Tracklet Calculation    &  150.0 &   43  &  179.2   &   -  &  329.2        \\  
    Projection Transceiver  &  150.0 &   13  &  54.2   & 316.7  &  520.8      \\  
    Projection Router       &  150.0 &    5  &   20.8   &  -  &  170.8        \\  
    Match Engine            &  150.0 &    6  &   25.0   &   -  &  175.0        \\  
    Match Calculator        &  150.0 &   16  &   66.7   &   -  &  216.7        \\  
    Match Transceiver       &  150.0 &   12  &  50.0   & 316.7  &  516.7      \\  
    Track Fit                    &  150.0 &   26  &  108.3   &   -  &  258.3        \\  
    Duplicate Removal     &  0.0 &   6  &   25.0   &   -  &  25.0        \\  
    Output Link             &    0.0 &    1  &    4.2   & 316.7  &  320.8        \\  
\hline 
    Total                   & 1500.0 &  139  &  579.2   & 1266.7  & \textbf{3345.8}    \\
    \hline
  \end{tabular}
\end{table}

\section{Tracklet System Performance} 
\label{sec:perf}

The demonstrator showed excellent agreement between the actual firmware results and the integer-based C++ emulation of the system. 
For single object events, the output tracks from the firmware have 100\% bitwise compatibility with the integer-based emulation.  
For busier events, for example top quark pair ($t\bar{t}$) events with an average PU = 200, the emulation and firmware tracks agree to better than 99\%. 
Because of this, the C++ emulation can safely be assumed to emulate well the demonstrator system. 

\begin{figure}[t!]
\centering
\includegraphics[width=7cm]{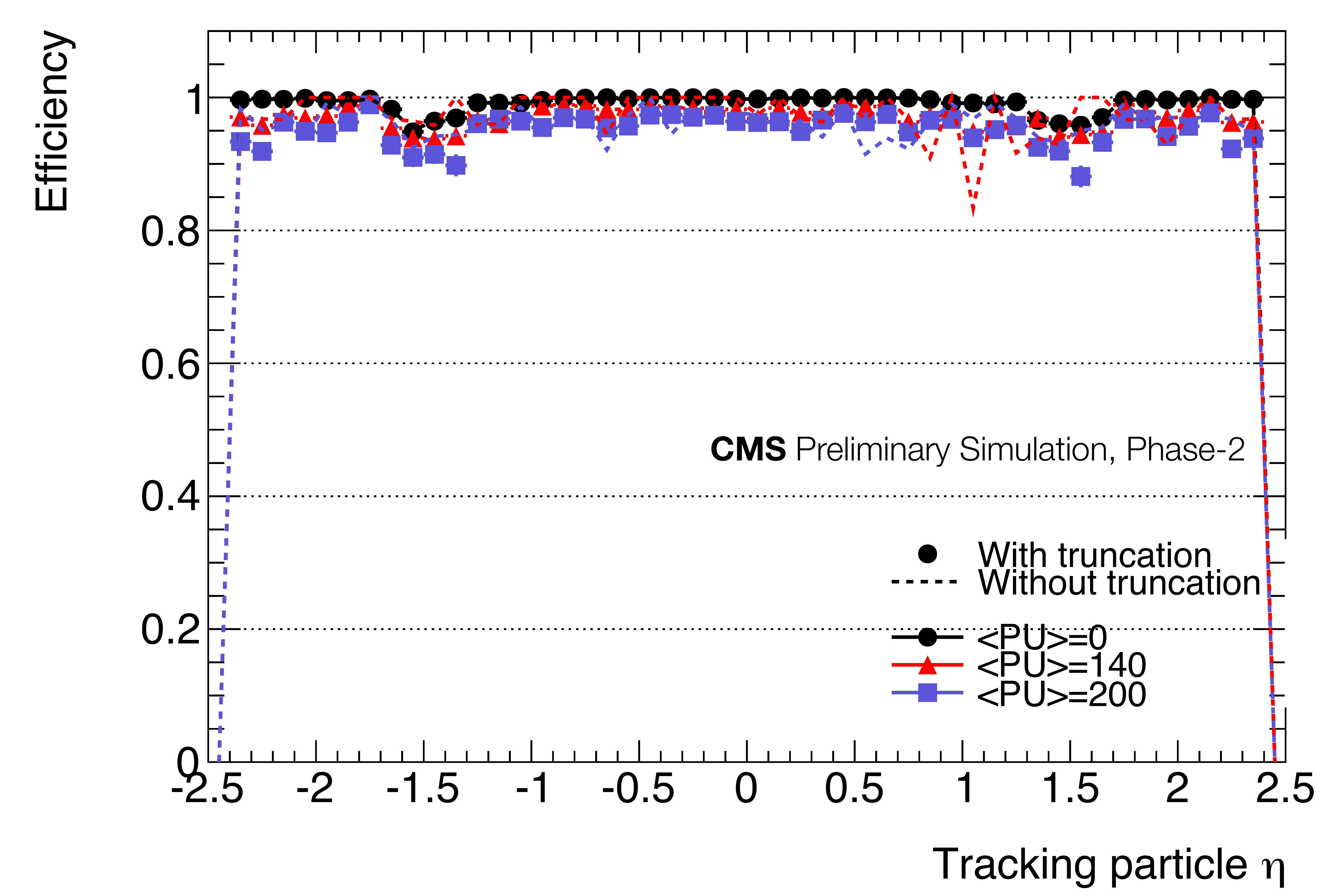}
\includegraphics[width=7cm]{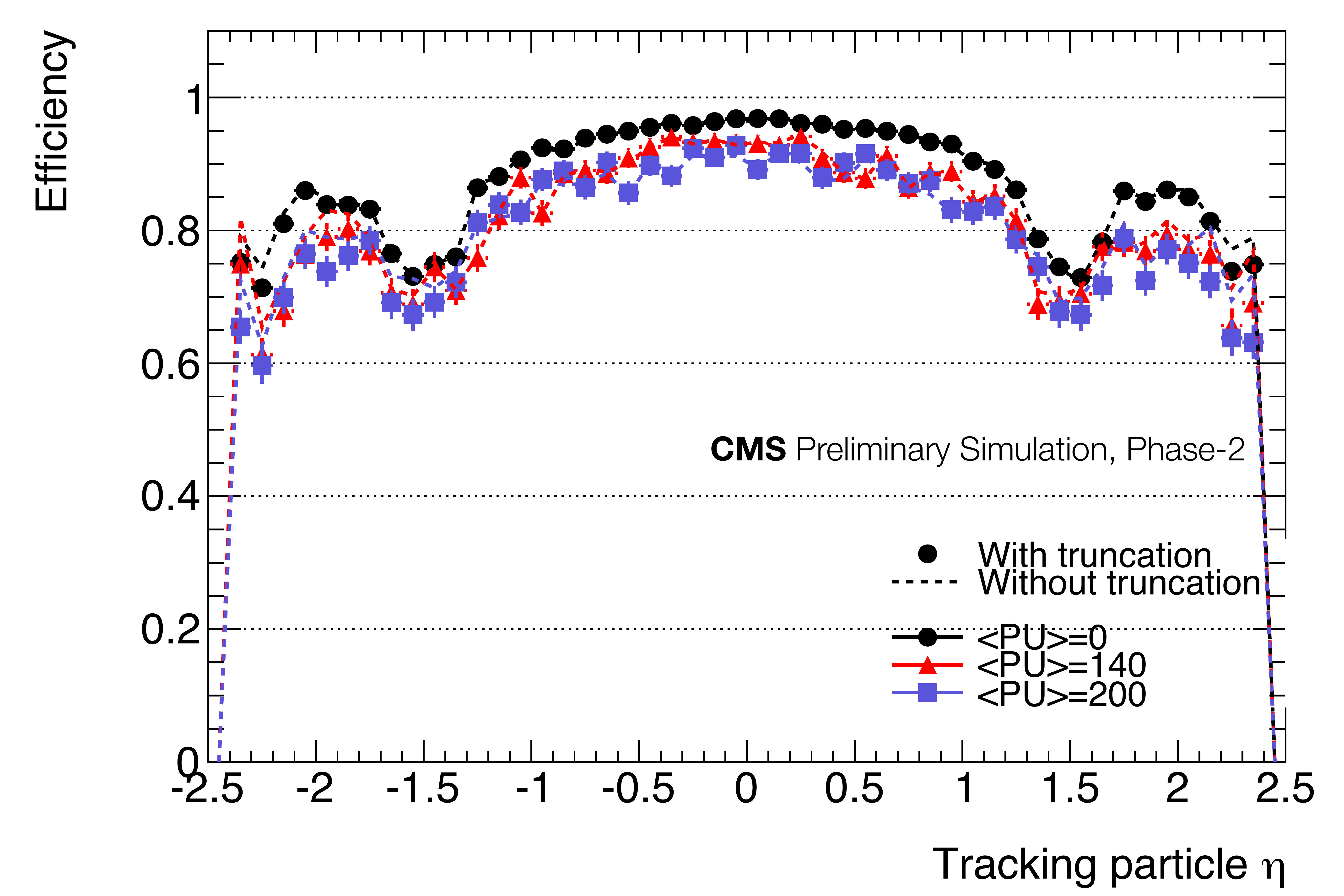}
\caption{ Efficiency as a function of $\eta$ for muons (left) and electrons (right). 
Results are shown for three different average pileup scenarios: 0 (black), 140 (red), and 200 (blue). 
The efficiency is shown with (filled points) and without (dashed lines) truncation included. }
\label{fig:efficiencies}
\end{figure}

The estimated performance of the tracklet algorithm is studied with the integer-based emulation of the algorithm. 
The efficiency of reconstructing the input tracks with all of their correct stubs as a 
function of $\eta$ for muon and electron events are shown in Figure \ref{fig:efficiencies}. 
For all of these objects, efficiencies are shown for several different average pileup conditions, $<$PU$>$ = 0, 140, and 200. 
The efficiency is computed as implemented in the demonstrator, i.e. it includes the effects of truncating data. 
Additionally the integer-based emulation provides an estimate of the efficiency without truncation effects. 
Efficiency with and without truncation effects are included in these plots.
Here it is seen that for all of these objects the data truncation has little effect on the track finding efficiency. 
The effect is minimal for two main reasons: (i) because of the large parallelization of the system, most of the modules are sparsely populated, 
 (ii) the different seeding combinations provide additional redundancy that can recover tracks that may otherwise be lost. 

\begin{figure}[ht]
\centering
\includegraphics[width=7cm]{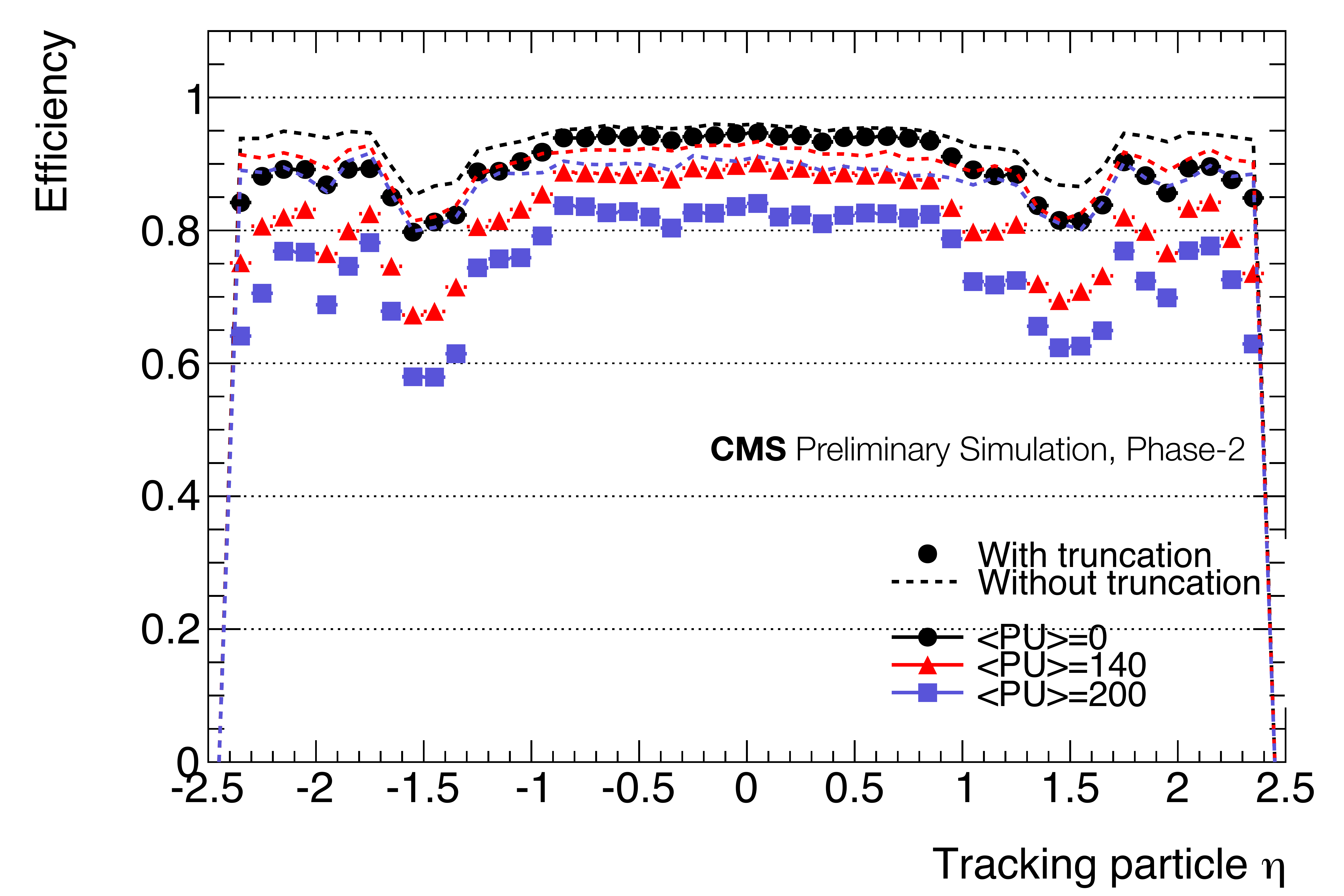}
\includegraphics[width=7cm]{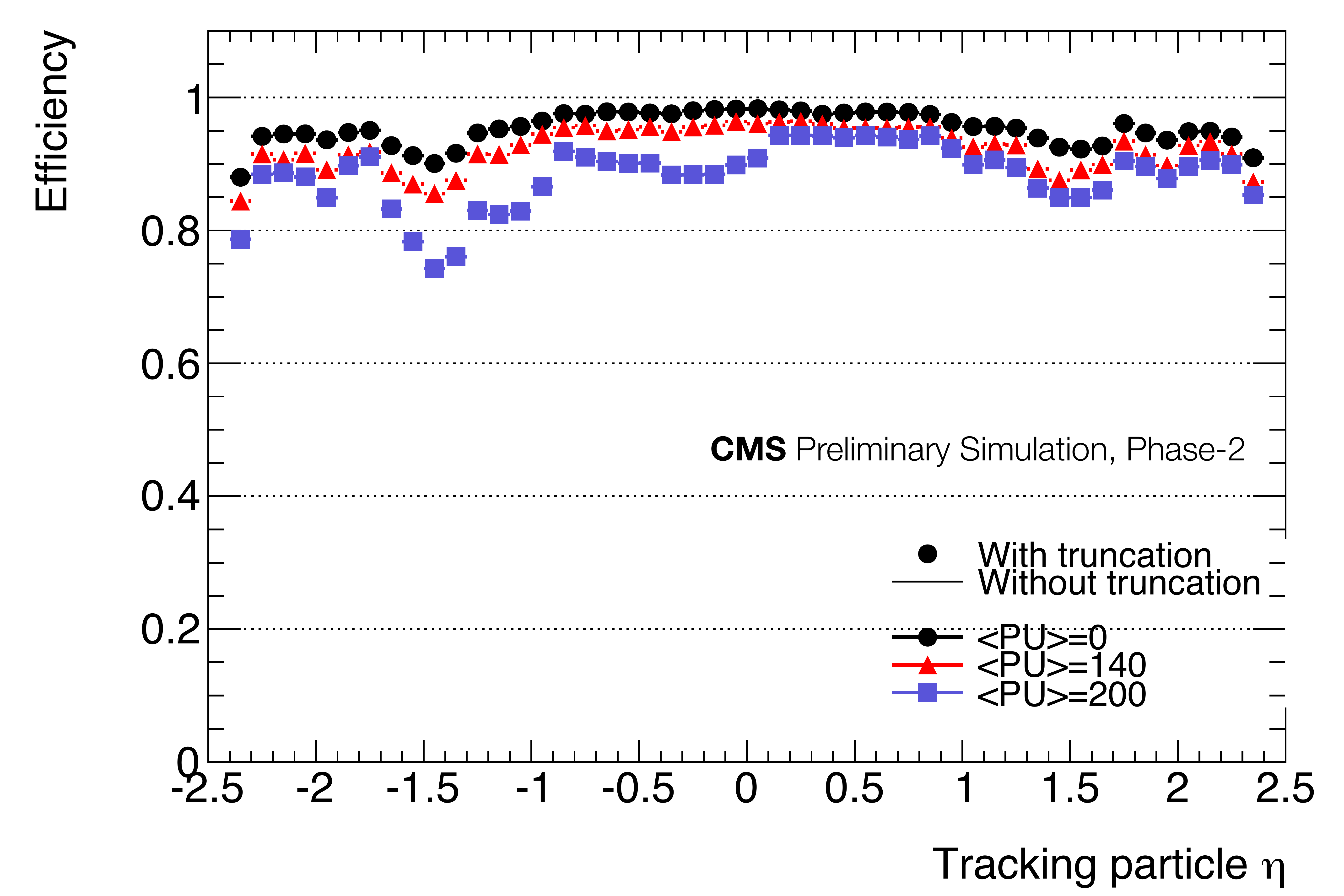}
\caption{ Efficiency as a function of $\eta$ for $t\bar{t}$ events with standard configuration (left) and after improved load balancing (right). 
Results are shown for three different average pileup scenarios: 0 (black), 140 (red), and 200 (blue). 
The efficiency is shown with and without truncation included as points and dashed lines, respectively. 
The asymmetry between positive-negative $\eta$ is introduced in the truncation and is currently under investigation. }
\label{fig:ttbar_eff}
\end{figure}

For very busy $t\bar{t}$ events there is a more drastic effect from truncation, as shown in Figure \ref{fig:ttbar_eff} (left).  
Each of the jets in these events are generally very dense (in $\phi$) and therefore are not well-split between virtual modules. 
In the combinatoric heavy stages of the algorithm there is not enough time to process all of the stubs, causing a drop in the efficiency. 
This can be fixed by better load-balancing. 
Changing the partitioning of virtual modules so that they are thinner in $\phi$ but span the entire $z$ of the detector alleviates these issues. 
With this new scheme, when there is a dense jet in an event, the stubs are spread over more virtual modules, meaning more stubs can be processed within the given latency and fewer stubs are lost due to truncation. 
The improvement in efficiency is shown in Figure \ref{fig:ttbar_eff} (right).  
Although this change increases by about 20\% the number of virtual modules per $\phi$ sector, there is no additional resource usage. 
The partitioning is thin enough to completely remove one of the lookup tables in the calculation, and reduces the number of memories needed downstream for storing the projections.
This improved load-balancing scheme is currently being implemented. 

\begin{figure}[ht]
\centering
\includegraphics[width=7cm]{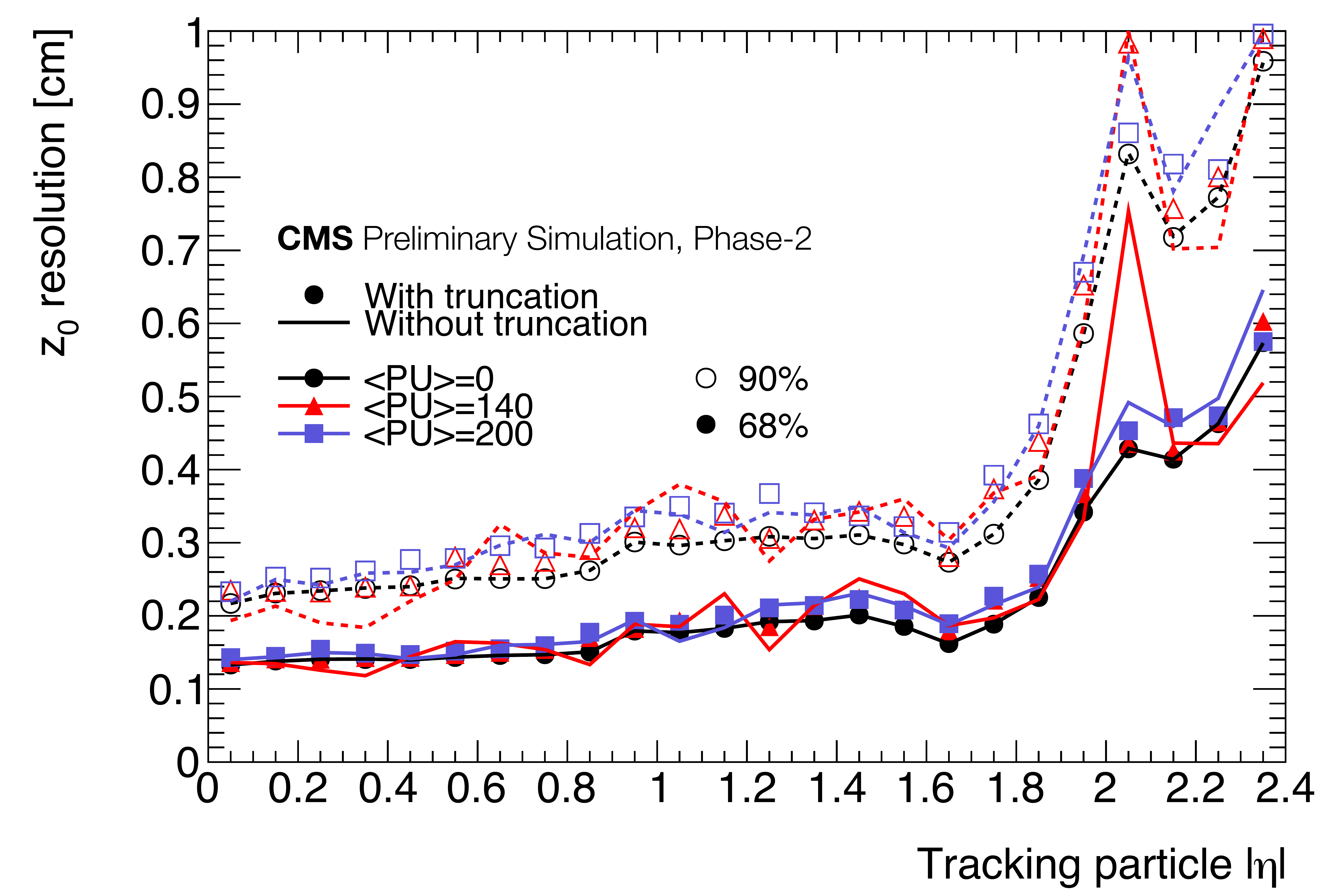}
\includegraphics[width=7cm]{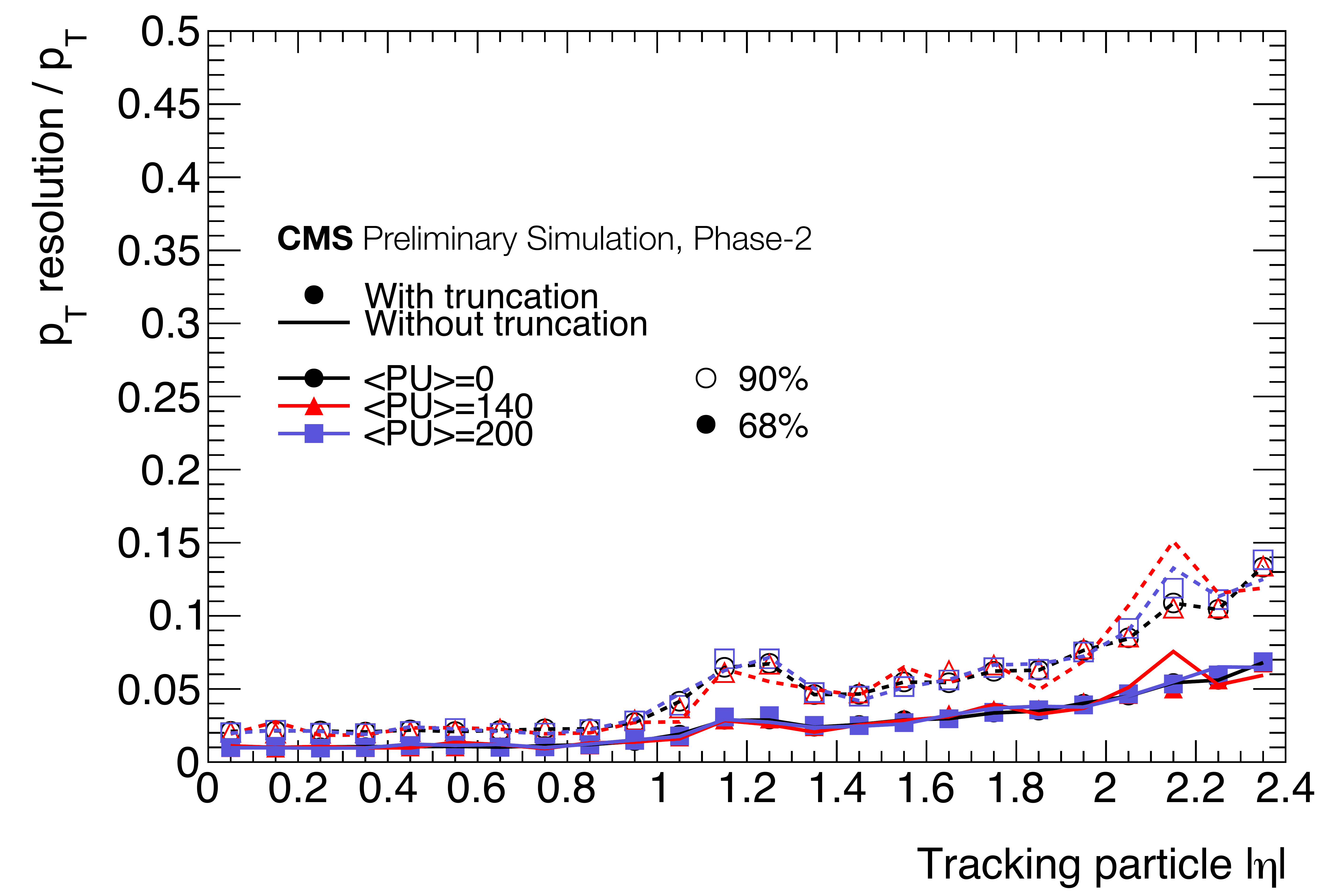}
\caption{ Resolution of $z_0$ (left) and relative $\pt$ (right) for the final tracks for muon events. 
Results are shown for three different average pileup scenarios: 0 (black), 140 (red), and 200 (blue). 
The resolution is shown with and without truncation included as points and dashed lines, respectively. 
}
\label{fig:resolutions}
\end{figure}

The tracklet algorithm achieves the resolution required of the L1 trigger system, specifically a $z_0$ resolution of 1-2 mm for $\eta < 1.9$ 
and a relative \pt resolution of less than 0.05 for almost the entire $\eta$ range as shown in Figure \ref{fig:resolutions}. 
The integer-based emulation achieves comparable resolution to that of a floating-point simulation of the tracklet algorithm \cite{ref:tracklet-louise}. 
Some improvements can still be made. Too few $\eta$ bins are used in the final track calculation lookup tables in the transition region between barrel and endcaps (the region with $\eta \approx 1.2$). This can be corrected, and will improve the resolution in that region. 
Additionally, the current version of the project uses too few bits for storing the stub $z$ position. This can also be corrected and will improve overall the track $z_0$ resolution shown here.

\begin{figure}[ht]
\centering
\includegraphics[width=7cm]{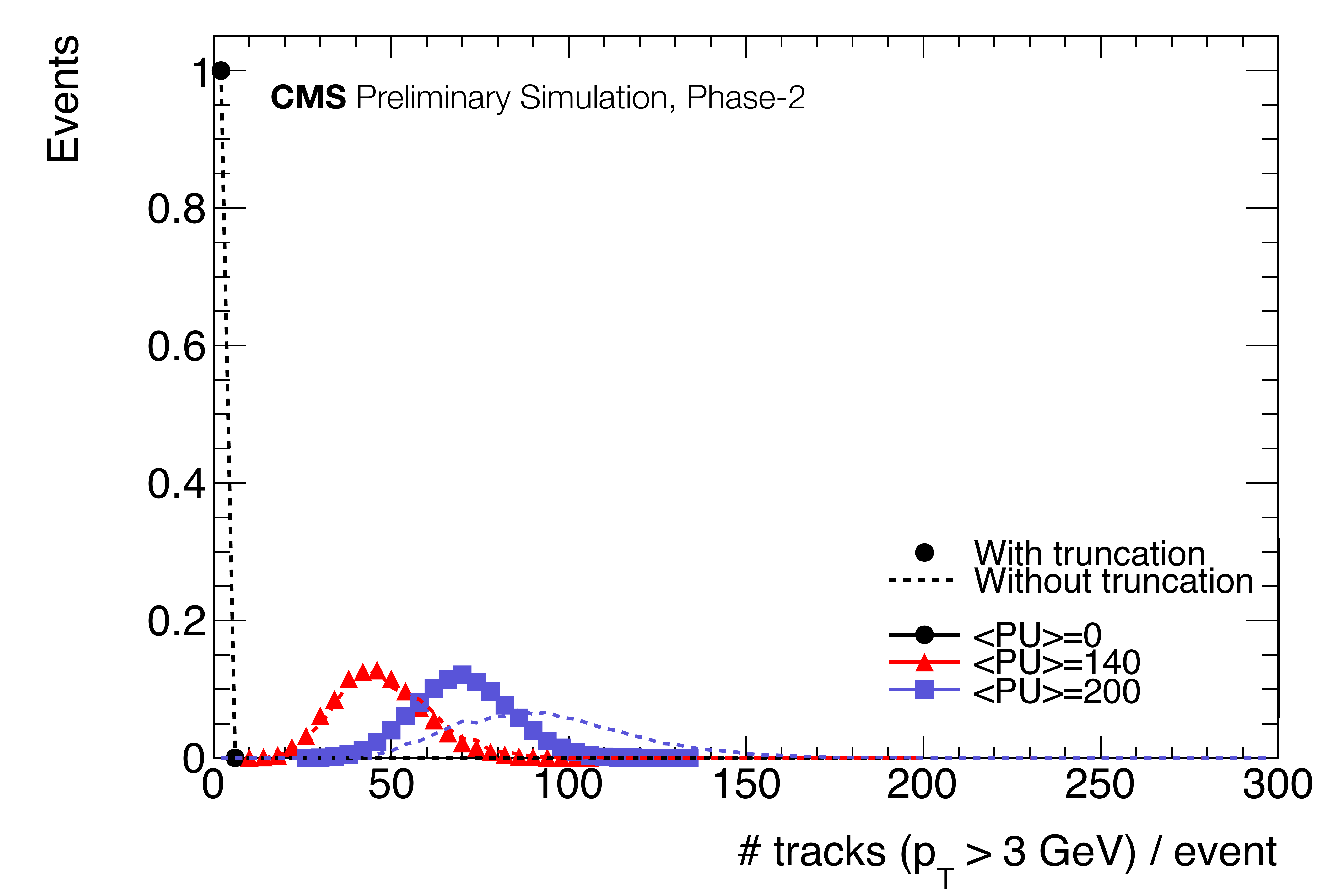}
\caption{Total number of tracks (with \pt $>$ 3 GeV) found for muons with three different average pileup scenarios: 0 (black), 140 (red), and 200 (blue). 
Results are shown with/without truncation included as filled points/dashed lines.}
\label{fig:rate}
\end{figure}

While achieving high efficiency and good resolution, the tracklet approach also achieves a relatively low fake rate (the rate of incorrectly reconstructing a track). 
The total number of tracks found for single muon events with different average pileup scenarios is shown in Figure \ref{fig:rate}. 
Here it can be seen that for a single muon without pileup, the tracklet algorithm almost always finds only one track. 
The fake rate in this scenario is extremely low.
With the addition of more pileup interactions, additional tracks are found.
More detailed studies on rates and track isolation for the tracklet approach can be found in Ref.~\cite{ref:tech-proposal}.

\section{Projections for a Full L1 Tracking System}
\label{sec:future}

Extrapolations can be made for a full tracklet track-trigger system to be used at CMS in the HL-LHC. 
Currently, almost an entire half $\phi$ sector (the $+z$ region of the detector) can be processed by a single Virtex-7 FPGA. 
Based on the resources available in this chip and what is used, it is anticipated that a full $\phi$ sector encompassing the full $z$ range of the detector will be able to be processed by a single future-generation FPGA. 
Actually the Virtex-7 FPGA has enough lookup table (LUT) resources to store the precomputed portions of the algorithm for the full $\phi$ sector.
Similarly this FPGA has the digital signal processing (DSP) resources needed to handle the relatively small number of calculations in the full algorithm.  
However, because the tracklet system relies largely on parallelization of the processing, each $\phi$ sector needs a lot of memory (in FPGAs these are fixed size block RAMs or BRAMs) to store intermediate steps in the calculation. 
The estimated resource usage for a full sector is shown in Table \ref{tab:fpgaresources}. 
As shown there, the needed resources, specifically the BRAM needs, surpass that of the current Virtex-7 FPGA.
However, a full $\phi$ sector processing needs fit comfortably within the resource allotments of the Xilinx Virtex-Ultrascale+ class of FPGAs, a few of which are shown in Table \ref{tab:fpgaresources}.  
The project resource needs can also be met by a Kintex-Ultrascale (specifically KU115) FPGA, 
if 16Gbps (instead of 25Gbps) links are sufficient for sending stubs from the detector to the L1 tracking system. 

\begin{table}[bth]
  \caption{FPGA resource utilization for a full $\phi$ sector based on the usage as reported by Xilinx's Vivado Design Suite for the implementation of a half $\phi$ sector. 
  The top line shows the needs for a full $\phi$ sector. The following lines show the fraction of
    the resource needs in Virtex-7 (top line) and Virtex
    Ultrascale+ (others) FPGAs. }
  \label{tab:fpgaresources}
  \centering
  \begin{tabular}{lllll}
    \hline
    &LUT Logic& LUT Memory& BRAM& DSP\\
    Full sector& 279733& 151191& 2721.5& 1818\\
    \hline\hline
    Virtex-7 690T& 65\%&87\%&185\%&51\%\\
    \hline
    VU3P     &32\%&81\%&85\%&80\%\\
    VU5P     & 21\%&53\%&58\%&52\%\\
    VU7P     &16\%&40\%&42\%&40\%\\
    VU9P     &11\%&27\%&28\%&27\%\\
    VU11P     &10\%&27\%&29\%&20\%\\
    VU13P     &  7\%&20\%&22\%&15\%\\
    \hline
  \end{tabular}
\end{table}

\section{Conclusions}
\label{sec:conclusion}

For the HL-LHC, CMS will require a new all-silicon tracker with trigger capabilities in order to achieve its physics goals. 
Tracking at the L1 trigger will provide greatly improved 
momentum measurements, enabling rate reducing triggers without increasing \pt thresholds. 
The tracklet approach is one of the proposed methods for performing track finding for the L1 trigger. 
The method is based on a road-search algorithm that is implemented on commercially available FPGA technology. 
The tracklet algorithm has been implemented as a floating-point simulation, integer-based emulation, and in an FPGA. 
Two projects are used to cover the entire $+z$ range of the detector. 
To demonstrate the system feasibility, a hardware demonstrator system based on Virtex-7 FPGAs was assembled and validated the algorithm's implementation. 
High performance, in terms of efficiency and resolution, is achieved. 
The demonstrator gives a measured latency for L1 track finding of 3.333 \mus. 
With little extrapolation, the system seems scalable to future FPGA technology.


\section*{Acknowledgements}
I would like to thank the CMS collaboration and contributors to this project at Cornell University, Rutgers University, Ohio State University, and University of Notre Dame.
I also wish to thank the Wisconsin University CMS group for their support of the CTP7 platform. 
This work was supported by the US National Science Foundation through the grant NSF-PHY-1307256.

\bibliography{CTD17_Tracklet_arXiv}
%
%
%
%

\end{document}